\documentclass[12pt]{iopart}

\usepackage{graphics} 
\usepackage{graphicx}
\usepackage{dcolumn}
\usepackage{bm}
\usepackage{subfigure}
\usepackage{multirow}
\usepackage{float} 
\begin{document}

\title{Scaling behaviours of the $p_{T}$ spectra for identified hadrons in $pp$ collisions}
\author{W. C. Zhang}
\address{School of Physics and Information Technology, Shaanxi Normal University, Xi'an 710119, People's Republic of China}
\eads{wenchao.zhang@snnu.edu.cn}
\begin{abstract}

  \noindent We extend the scaling behaviour observed in the inclusive charged hadron transverse momentum ($p_{T}$) distributions to the $p_{T}$ spectra of pions, kaons and protons produced in proton-proton ($pp$) collisions with center of mass energies ($\sqrt{s}$ ) at  0.9, 2.76 and 7 TeV. This scaling behaviour arises when a linear transformation, $p_{T} \rightarrow p_{T}/K$,  is applied on the pion, kaon or proton $p_{T}$ spectra. The scaling parameter $K$ depends on $\sqrt{s}$ and is determined by a new method, the quality factor method, which does not rely on the shape of the scaling function. We argue that the pions, kaons and protons originate from different distributions of clusters which are formed by strings overlapping, and the scaling behaviours of these identified particles $p_{T}$ spectra could be understood with the colour string percolation model in a quantitative way simultaneously.

\end{abstract}
\pacs{13.85.Ni, 13.87.Fh}

\maketitle

\section{Introduction}\label{Intro}
One of the most important observations in high energy collisions is the $p_{T}$ spectra for different species of final state particles. From the spectra, we can learn a lot about the regularities of the particle productions.

In many studies, searching for a scaling behaviour of the $p_{T}$ spectra is helpful to reveal these regularities. In \cite{pion_spectrum}, the authors showed a scaling behaviour in the pion $p_{T}$ spectra with different collision centralities at midrapidity in Au+Au collisions at the Relativistic Heavy Ion Collider (RHIC). This scaling behaviour was extended to noncentral regions in Au+Au and d+Au collisions \cite{non_central_collision}. Similar scaling behaviours were found in the proton and anti-proton $p_{T}$ spectra with different collision centralities at midrapidity in Au+Au collisions at RHIC \cite{proton_antiproton_spectra}.

Recently, we observed a scaling behaviour in the $p_{T}$ spectra of inclusive charged hadrons in 
proton-proton (proton-antiproton) collisions at $\sqrt{s}$ = 0.9, 2.76 and 7 (0.63, 1.8 and 1.96) TeV when a linear transformation, $p_{T} \rightarrow z=p_{T}/K$, was applied on these spectra \cite{pp_ppbar_collisions}. Here the scaling parameter $K$ depends on the collision energy $\sqrt{s}$. This scaling behaviour was explained by the colour string percolation model in a qualitative way. In this paper, we will extend this scaling behaviour to the $p_{T}$ spectra of pions, kaons and protons produced in $pp$ collisions at  0.9, 2.76 and 7 TeV. A new method, the quality factor method \cite{QF_1,QF_2}, will be adopted in the searching for the scaling parameter $K$. Compared with the method utilized in \cite{pp_ppbar_collisions}, this method does not rely on the shape of the scaling function, thus it is more robust. We will argue that the pions, kaons and protons originate from different distributions of clusters which are formed by strings overlapping, and the colour string percolation model can describe the scaling behaviours of the identified $p_{T}$ spectra not only in a qualitative way but also in a quantitative way simultaneously. As a similar geometrical scaling behaviour was suggested when the $p_{T}$ spectra of identified particles in $pp$ collisions at 0.9, 2.76 and 7 TeV were presented in terms of a scaling variable $\tau=p_{T}/Q_{0}(p_{T}/\sqrt{s})^{\lambda/2}$, where $Q_{0}=1$ GeV/c and $\lambda$ is around 0.27 \cite {geometrical_scaling}, we would like to compare the scaling behaviours presented in $z$ and $\tau$.

The organization of the paper is as follows. In section \ref{sec:method}, we will illustrate the procedure to search for the scaling behaviours in the $p_{T}$ spectra of identified particles. In section \ref{sec:scaling_behaviour}, the scaling behaviours of pions, kaons and protons will be described.  Section \ref{sec:comp} shows the comparison between the scaling behaviours presented in the variables $z$ and $\tau$. In section \ref{sec:mechanism},  we will use the colour string percolation model to explain the scaling behaviours of the identified particles in a quantitative way. Finally, the conclusion is made in section \ref{sec:conclusion}.

\section{Method to search for the scaling behaviours} \label{sec:method}

As done in \cite{pp_ppbar_collisions}, we will search for the scaling behaviours of the identified pion, kaon and proton $p_{T}$ spectra with the following steps. Let us take the pion $p_{T}$ spectra as an example, we define a scaling variable, $z=p_{T}/K$, and a scaled $p_{T}$ spectra, $\Phi(z)=A\cdot(2\pi p_{T})^{-1}d^{2}N/dp_{T}dy|_{p_{T}=Kz}$. Here the parameters $K$ and $A$ depend on the collision energy. By choosing proper $K$ and $A$, the data points of the pion $p_{T}$ spectra in $pp$ collisions at 0.9, 2.76 and 7 TeV will migrate into one curve. As a convention,  we usually set both $K$ and $A$ of the highest energy collisions to be 1. However, the $p_{T}$ coverages of the pion spectra at 0.9 and 7 TeV are much smaller than the coverage at 2.76 TeV. In order to make the scaling function $\Phi(z)$ to describe the pion spectra in the large $p_{T}$ region faithfully, we choose both $K$ and $A$ for the collisions at 2.76 TeV as 1.  With different choices of $K$ and $A$, the scaling functions $\Phi(z)$ are different. In order to get rid of this arbitrariness, we introduce another scaling variable, $u=z/\langle z \rangle=p_{T}/\langle p_{T} \rangle$, and the corresponding normalized scaling function $\Psi(u)=\langle z \rangle^{2}\Phi(\langle z \rangle u)\big/\int^{\infty}_{0}\Phi(z)zdz$. Here $\langle z \rangle=\int^{\infty}_{0}z\Phi(z)zdz\big/\int^{\infty}_{0}\Phi(z)zdz$. The ways to search for the scaling behaviours in the $p_{T}$ spectra of kaons and protons are identical to the one for pions.

\section{Scaling behaviours of identified pions, kaons and protons }\label{sec:scaling_behaviour}

In this paper, the $p_{T}$ spectra of identified pions, kaons and protons in $pp$ collisions at 0.9, 2.76 and 7 TeV were published by the ALICE collaboration \cite{pt_spectra_0_9, pt_spectra_2_76, pt_spectra_7_0}. Here the identified pion, kaon and proton $p_{T}$ spectra refer to the spectra of $(\pi^{+}+\pi^{-})/2$, $(K^{+}+K^{-})/2$ and $(p+\bar{p})/2$. The data at 2.76 TeV cover a pion $p_{T}$ range of 0.1-20.0 GeV/c, which is much wider than the ones at 0.9 and 7 TeV, 0.1-2.6 GeV/c and 0.1-3.0 GeV/c. Similar results are obtained in the comparison among the kaon or proton $p_{T}$ ranges at 0.9, 2.76 and 7 TeV. As described in section \ref{sec:method}, both the parameters $K$ and $A$ at 2.76 TeV are set as 1,  thus the scaling function $\Phi(z)$ is nothing but the $p_{T}$ spectra of pions, kaons or protons at this energy. In \cite{pp_ppbar_collisions}, $\Phi(z)$ is written in the Tsallis form, in which the mass effect of the charged hadron is ignored. In this work, since the threshold of the $p_{T}$ range is 0.1 (0.1 and 0.1) GeV/c at 2.76 TeV \cite{pt_spectra_7_0}, which is below the mass of pions (kaons and protons) 0.14 (0.494 and 0.938) GeV/c$^{2}$ \cite{pdg}, we write the scaling function $\Phi(z)$ for pions, kaons or protons with the modified Tsallis form
\begin{eqnarray}
\Phi(z)=C_{q}\left[1-(1-q)\frac{\sqrt{m^{2}+z^{2}}-m}{z_{0}}\right]^{\frac{1}{1-q}},
\label{eq:phi_z_pt_spectrum_pp}
\end{eqnarray}
where $C_{q}$, $q$ and $z_{0}$ are free parameters, $m$ is the mass of the particle species, and $1-q$ is a measure of the non-extensivity. These free parameters are given by fitting equation \eref{eq:phi_z_pt_spectrum_pp} to the $p_{T}$ spectra of identified species at 2.76 TeV with the least $\chi^{2}$s method. The square root of the sum of the statistical and systematic uncertainties of the data points has been taken into account in these fits. \Tref{tab:id_particles_fit_parameters} lists the parameters $C_{q}$, $q$, $z_{0}$ as well as their uncertainties returned by the fits. The last line of this table shows the $\chi^2$s per degrees of freedom (dof), named reduced $\chi^{2}$s, for these fits.

\begin{table}[H]
\caption{\label{tab:id_particles_fit_parameters} The parameters $C_{q}$, $q$ and $z_{0}$ of the scaling functions $\Phi(z)$ for pions, kaons and protons. The uncertainties quoted are due to the statistical plus systematic errors of the data points added in quadrature. The last line shows the reduced $\chi^{2}$s for the fits on the $p_{T}$ spectra of identified particles at  2.76 TeV. }
\begin{center}
\begin{tabular}{@{}cccc}
\hline  
\textrm{\ }&
\textrm{Pions}&
\textrm{Kaons}&
\textrm{Protons}
\\
\hline  
$C_{q}$ & 6.01$\pm$0.08 & 0.214$\pm$0.002& 0.0546$\pm$0.0008\\
$q$&1.1416$\pm$0.0007 & 1.1402$\pm$0.0004& 1.115$\pm$0.002\\
$z_{0}$ (GeV/c)& 0.1308$\pm$0.0008 & 0.193$\pm$0.001& 0.220$\pm$0.002\\
$\chi^{2}$/dof& 0.49 & 0.23& 0.40\\
\hline  
\end{tabular}
\end{center}
\end{table}

In \cite{pp_ppbar_collisions}, for $pp$ collisions, the scaling parameters $K$ and $A$ at 0.9 and 2.76 TeV are determined by fitting the scaled Tsallis distribution in equation (2) of that reference to the $p_{T}$ spectra at these two energies. The quality of the fit heavily depends on the parameters in the scaled Tsallis distribution, $C_{q}$, $q$ and $z_{0}$, which are fixed to the values obtained at 7 TeV. Thus $K$ and $A$ at 0.9 and 2.76 TeV rely on the shape of the scaling function in equation (1) of the reference. In this paper, in order to eliminate this reliance, we adopt the quality factor (QF) method to determine $K$ and $A$ for pions, kaons and protons at 0.9 and 7 TeV. In this method, the QF is defined in terms of a set of data points $(u^{i}, v^{i})$  \cite{QF_1,QF_2}
\begin{eqnarray}
QF(K,A)=\left[\sum_{i}\frac{(v^{i}-v^{i-1})^{2}}{(u^{i}-u^{i-1})^{2}+\varepsilon^{2}}\right]^{-1},
\label{eq:QF_definition}
\end{eqnarray}
where $u^{i}=p_{T}^{i}/K$, $v^{i}=\textrm{log}(A\cdot(2\pi p^{i}_{T})^{-1}d^{2}N^{i}/dp^{i}_{T}dy^{i})$, the small constant $\varepsilon$ is taken as 0.01 and is utilized to keep the sum being finite when two points have the same $u$ value. Before entering the QF formula, $(u^{i}, v^{i})$ has been rescaled so that  $0 \leq u^{i},v^{i} \leq 1$, and $u^{i}$ are ordered. Obviously, two successive data points being close in $u$ and far in $v$ will give a large contribution to the sum in equation \eref{eq:QF_definition}. As a result, a set of data points with a small sum (thus a large QF) is expected to lie close to a unique curve. For the scaling parameters at 0.9 (7) TeV, we utilize the data points at 0.9 (7) and 2.76 TeV to determine them. The best set of ($K$,  $A$) for 0.9 (7) TeV is chosen to be the one which globally maximizes the QF. \Tref{tab:id_particles_a_k_parameters} tabulates $K$ and $A$ for pions, kaons and protons in $pp$ collisions at 0.9, 2.76 and 7 TeV. In equation \eref{eq:QF_definition}, the errors of data points are not taken into account. As described in \cite{QF_1}, these errors can be introduced into the QF in the following way: 
\begin{eqnarray}
QF(K,A)=\left[\sum_{i}\frac{(v^{i}-v^{i-1})^{2}\rho^{i}\rho^{i-1}}{(u^{i}-u^{i-1})^{2}+\varepsilon^{2}}\right]^{-1},
\label{eq:QF_definition_1}
\end{eqnarray}
where $\rho^{i}$ is the statistical uncertainty of data point $i$. Obviously, with this definition, the statistical scatter of different data points now have been taken into account in the QF. By maximizing the QF in equation \eref{eq:QF_definition_1}, we can get another set of ($K, A$). We take the difference between the two sets of ($K, A$) returned by the maximization of the QFs in equations \eref{eq:QF_definition} and \eref{eq:QF_definition_1} as the uncertainties of ($K, A$).



\begin{table}[H]
  \caption{\label{tab:id_particles_a_k_parameters} $K$ and $A$ for pions, kaons and protons at 0.9, 2.76 and 7 TeV. The errors quoted are taken from the difference between the two sets of ($K, A$) returned by the maximization of the QFs in equations \eref{eq:QF_definition} and \eref{eq:QF_definition_1}.}
\begin{center}
\begin{tabular}{@{}ccccc}
 \hline
 \textrm{\ }& 
\textrm{$\sqrt{s}$ (TeV)}&
\textrm{$K$}&
\textrm{$A$}\\
\hline  
 \textrm{\ }& 0.9 & 0.93$\pm$0.04& 1.11$\pm$0.14\\
 \textrm{Pions}& 2.76 &1 & 1\\
 \textrm{\ }& 7 & 1.06$\pm$0.02 & 0.89$\pm$0.07\\
 \hline
 \textrm{\ }& 0.9 & 0.913$\pm$0.007 & 1.05$\pm$0.01\\
 \textrm{Kaons}& 2.76& 1&1\\
 \textrm{\ }& 7 & 1.14$\pm$0.04 & 1.05$\pm$0.14\\
 \hline
\textrm{\ }& 0.9 & 0.926$\pm$0.001 & 1.08$\pm$0.02\\
 \textrm{Protons}& 2.76 &1 & 1\\
 \textrm{\ }& 7 & 1.108$\pm$0.006 & 1.04$\pm$0.02\\
\hline
\end{tabular}
\end{center}
\end{table}

\begin{figure}[h]
\centering
\includegraphics[scale=0.18]{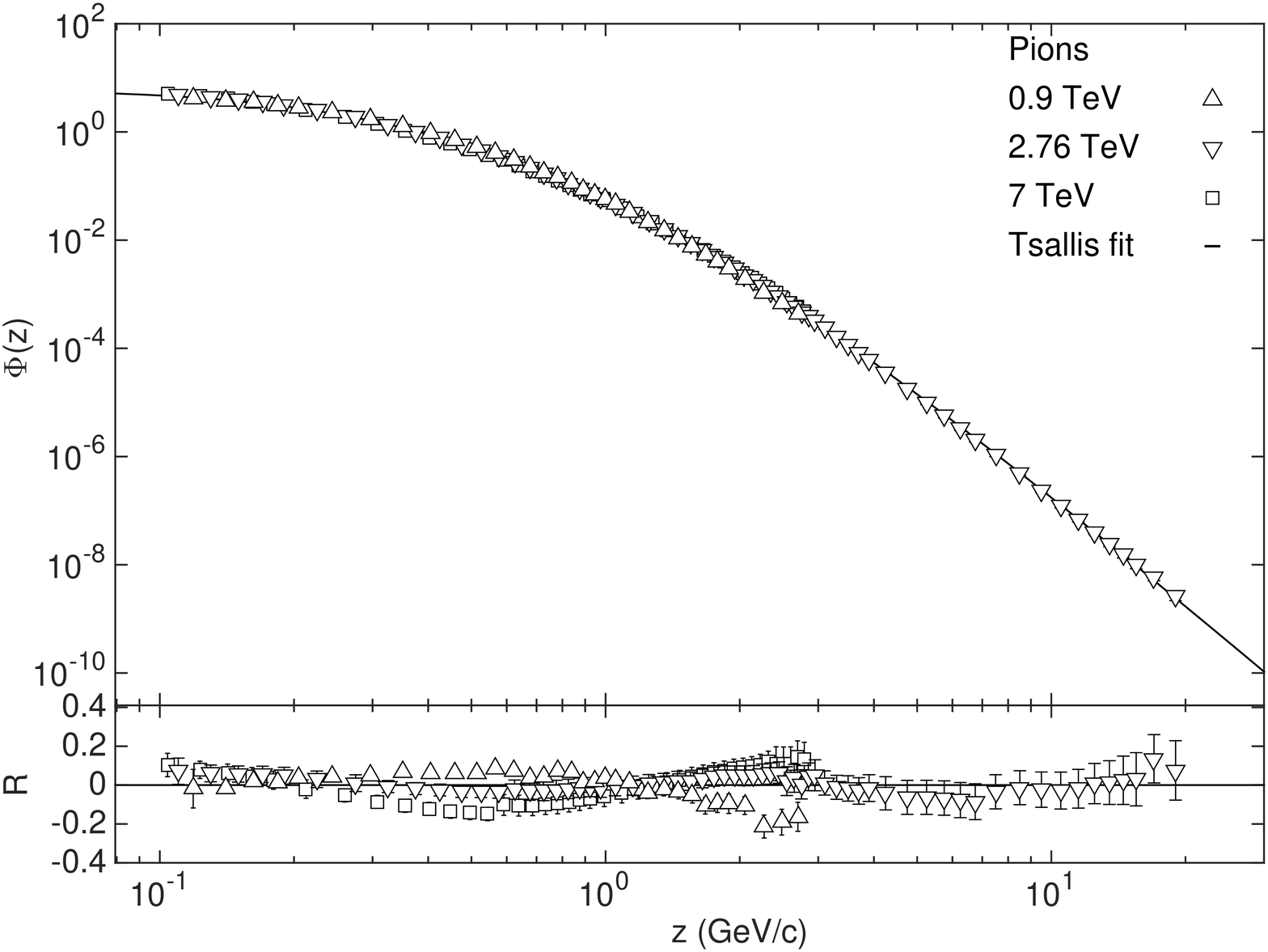} 
\caption{\label{fig:pi_z_plus_ratio_log_2_76_TeV_QF}Upper panel: the scaling behaviour of the pion $p_{T}$ spectra presented in $z$ for $pp$ collisions at 0.9, 2.76 and 7 TeV. The solid curve is described by equation \eref{eq:phi_z_pt_spectrum_pp} with parameters in the second column of \tref{tab:id_particles_fit_parameters}. The data points are taken from \cite{pt_spectra_0_9, pt_spectra_2_76, pt_spectra_7_0}. Lower Panel: the $R$ distribution as a function of $z$ for pions.}
\end{figure}

Since the scaling parameters $K$ and $A$ at 0.9 and 7 TeV now are obtained with the QF method, we could alternatively determine the scaling function $\Phi(z)$ by fitting equation \eref{eq:phi_z_pt_spectrum_pp} to the combined data points at 0.9, 2.76 and 7 TeV, rather than to the  points only at 2.76 TeV. However, the parameters $C_{q}$, $q$ and $z_{0}$ returned by this fit are consistent with the ones in \tref{tab:id_particles_fit_parameters} when considering their uncertainties. Thus we will use these three parameters in \tref{tab:id_particles_fit_parameters} throughout this work. With $K$ and $A$ in \tref{tab:id_particles_a_k_parameters}, we plot the scaled pion $p_{T}$ spectra at 0.9, 2.76 and 7 TeV in terms of the scaling variable $z$ in the upper panel of \fref{fig:pi_z_plus_ratio_log_2_76_TeV_QF}. In the log scale, we observe that all data points at different energies now are shifted to the same curve within error bars. This curve is described by the pion scaling function $\Phi(z)$ in equation \eref{eq:phi_z_pt_spectrum_pp} with parameters in the second column of  \tref{tab:id_particles_fit_parameters}. In order to see the agreement between the experimental data and the fitted results, we evaluate the ratios, $R=\rm (data-fitted)/data$, at 0.9, 2.76 and 7 TeV.  The lower panel of \fref{fig:pi_z_plus_ratio_log_2_76_TeV_QF} shows the $R$ distribution as a function of $z$. It is obvious that the $R$ values of all data points are in the range between -0.2 and 0.2, which implies that the data and fitted curve agreement is within 20$\%$. This agreement roughly corresponds to the systematic error on $R$ and the accuracy of the fit. If we take into account the systematic uncertainties on $R$, then the accuracy of the fit is around 10$\%$. Considering the fact that the data in the pion $p_{T}$ spectra cover about 10 orders of magnitude, the fit performed on the pion $p_{T}$ spectra is good.

\begin{figure}[h]
\centering
\includegraphics[scale=0.18]{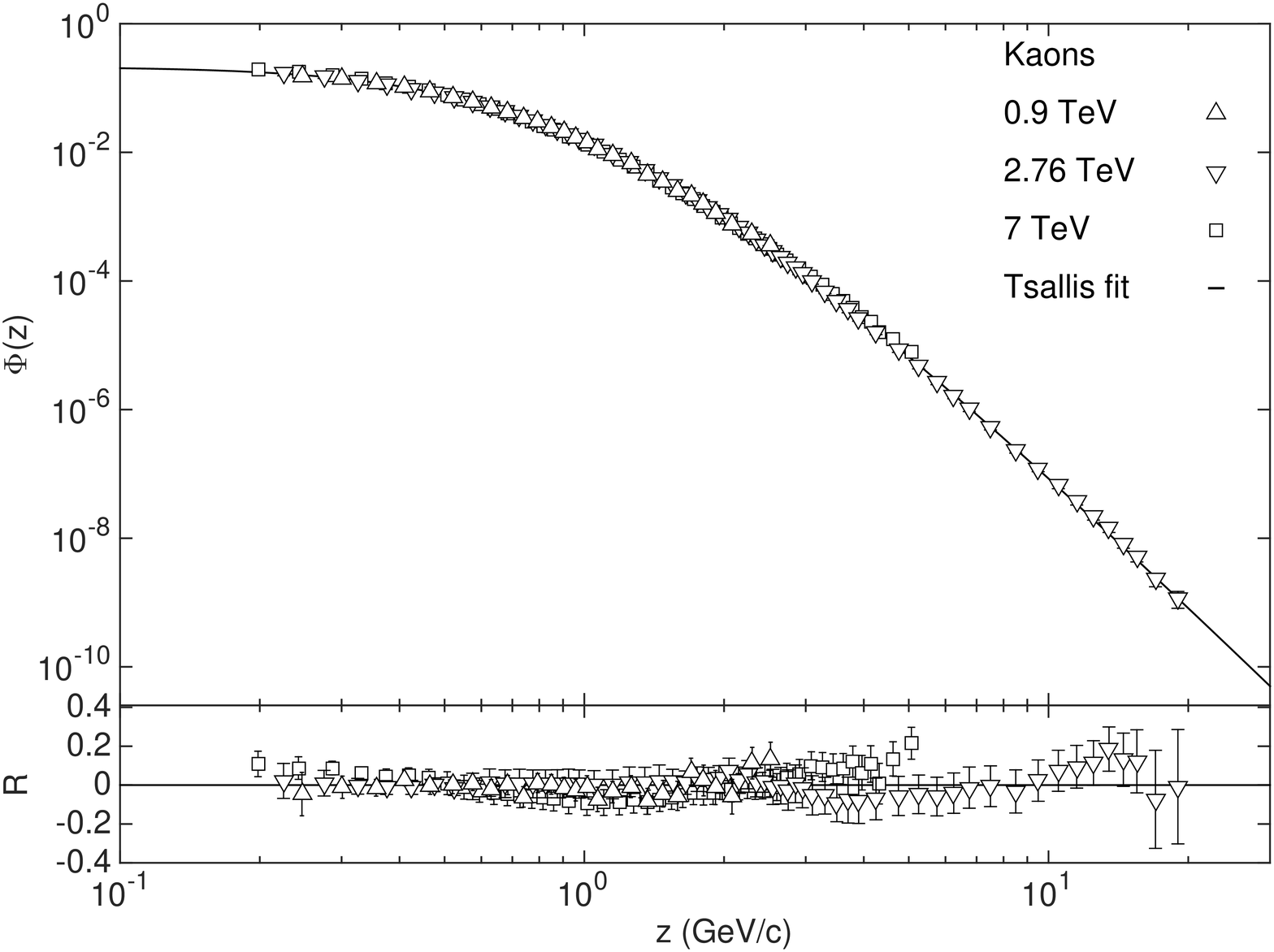} 
\caption{\label{fig:k_z_plus_ratio_log_2_76_TeV_QF}Upper panel: the scaling behaviour of the kaon $p_{T}$ spectra presented in $z$ for $pp$ collisions at  0.9, 2.76 and 7 TeV. The solid curve is described by equation \eref{eq:phi_z_pt_spectrum_pp} with parameters in the third column of  \tref{tab:id_particles_fit_parameters}. The data points are taken from \cite{pt_spectra_0_9, pt_spectra_2_76, pt_spectra_7_0}. Lower Panel: the $R$ distribution as a function of $z$ for kaons.}
\end{figure}

\begin{figure}[h]
\centering
\includegraphics[scale=0.18]{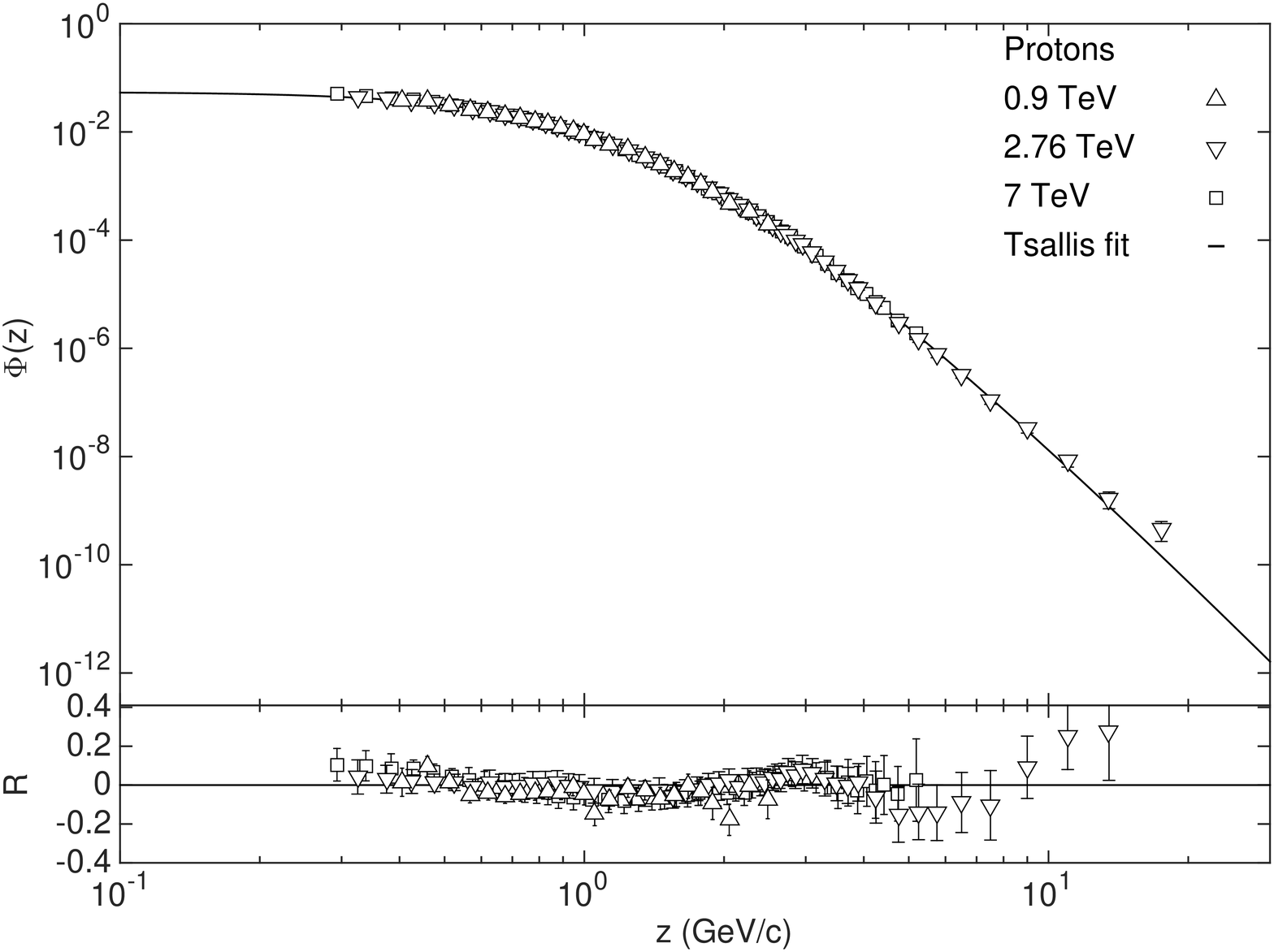} 
\caption{\label{fig:p_z_plus_ratio_log_2_76_TeV_QF}Upper panel: the scaling behaviour of the proton $p_{T}$ spectra presented in $z$ for $pp$ collisions at  0.9, 2.76 and 7 TeV. The solid curve is described by equation \eref{eq:phi_z_pt_spectrum_pp} with parameters in the fourth column of \tref{tab:id_particles_fit_parameters}. The data points are taken from \cite{pt_spectra_0_9, pt_spectra_2_76, pt_spectra_7_0}. Lower Panel: the $R$ distribution as a function of $z$ for protons. The $R$ value for the last data point at 7 TeV is around 0.7 and not shown in the lower panel.}
\end{figure}

The scaling behaviours of the kaon and proton $p_{T}$ spectra at 0.9, 2.76 and 7 TeV are presented in the upper panels of figures \ref{fig:k_z_plus_ratio_log_2_76_TeV_QF} and \ref{fig:p_z_plus_ratio_log_2_76_TeV_QF}. As done for pions, we show the $R$ distributions for kaons and protons in the lower panels of these two figures. For kaons, except for the last data point at 7 TeV, all the data points and the fitted curve agree within 20$\%$. For protons, the data points are consistent with the curve within 20$\%$ when $z$ is smaller than 10 GeV/c. When $z>10$ GeV/c, the yield and experimental errors for protons at 2.76 TeV have the same order. Thus the values of $R$ are large in this high $p_{T}$ region.


So far we have seen that the $p_{T}$ spectra of pions, kaons or protons at 0.9, 2.76 and 7 TeV indeed exhibit a scaling behaviour when these spectra are expressed in terms of $z$. The scaling function $\Phi(z)$ could be written as the modified Tsallis distribution in equation \eref{eq:phi_z_pt_spectrum_pp} phenomenologically.  As described in \cite{q_parameter_explanation}, the $q$ parameter of the Tsallis distribution for the original $p_{T}$ spectra of identified particles depends on the collision energy.  In this work, the conclusion drawn from the scaling behaviours of identified particles $p_{T}$ spectra is that the $q$ parameters of the scaled (not the original) $p_{T}$ spectra at 0.9 and 7 TeV are the same as the one at 2.76 TeV. Thus this conclusion is not in contradiction with the result from \cite{q_parameter_explanation}. Since the scaling function $\Phi(z)$ depends on the choice of the scaling parameters $K$ and $A$ at 2.76 TeV, we would like to eliminate it by replacing $z$ with $u=z/\langle z \rangle$. The $\langle z \rangle$ values for the pion, kaon and proton $p_{T}$ spectra are 0.442$\pm$0.007, 0.701$\pm$0.008 and 0.83$\pm$0.02 GeV/c, where the errors originate from the uncertainties of $C_{q}$, $q$ and $z_{0}$. With the substitution of $\langle z \rangle$ and $\Phi(z)$ into $\Psi(u)$ defined in section \ref{sec:method}, we can get the normalized scaling functions for pions, kaons and protons as 
\begin{eqnarray}
\Psi(u)=C^{'}_{q}\left[1-(1-q^{'})\frac{\sqrt{(m^{'})^{2}+u^{2}}-m^{'}}{u_{0}}\right]^{\frac{1}{1-q^{'}}},
\label{eq:psi_u_pt_spectrum_pp}
\end{eqnarray}
where $C^{'}_{q}=\langle z \rangle^{2}C_{q}/\int^{\infty}_{0}\Phi(z)zdz$, $q^{'}=q$, $u_{0}=z_{0}/\langle z \rangle$ and $m^{'}=m/\langle z \rangle$. \Tref{tab:id_particles_normalized_parameters} shows the values of $C^{'}_{q}$, $q^{'}$ and $u_{0}$ for these identified particles.

\begin{table}[H]
\caption{\label{tab:id_particles_normalized_parameters} The parameters $C^{'}_{q}$, $q^{'}$ and $u_{0}$ in the normalized scaling functions $\Psi(u)$ for pions, kaons and protons. The uncertainties quoted are due to the errors of parameters $C_{q}$, $q$ and $z_{0}$ in \tref{tab:id_particles_fit_parameters}. }
\begin{center}
\begin{tabular}{@{}cccc}
\hline  
\textrm{\ }&
\textrm{Pions}&
\textrm{Kaons}&
\textrm{Protons}
\\
\hline  
$C^{'}_{q}$ & 3.97$\pm$0.03& 2.86$\pm$0.03& 2.28$\pm$0.03\\
$q^{'}$&1.1416$\pm$0.0007 & 1.1402$\pm$0.0004& 1.115$\pm$0.002\\
$u_{0}$ & 0.296$\pm$0.009 & 0.275$\pm$0.004& 0.26$\pm$0.01\\
\hline  
\end{tabular}
\end{center}
\end{table}

\begin{figure}[h]
\centering
\includegraphics[scale=0.18]{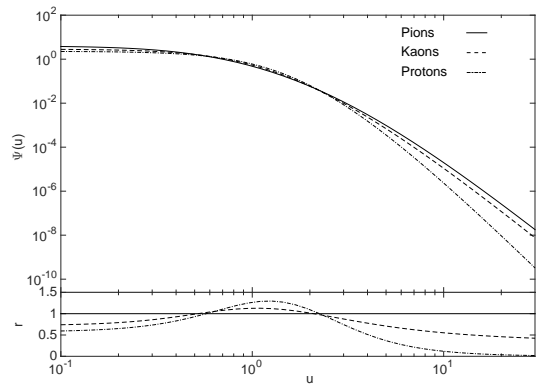} 
\caption{\label{fig:comp_u_pi_k_p_tsallis_diff}Upper panel: the normalized scaling functions for pions, kaons and protons. Lower Panel: the distribution of the ratio between the normalized scaling functions of kaons (protons) and pions.}
\end{figure}

Now we would like to explore the difference among the normalized scaling functions $\Psi_{\rm Pions}(u)$, $\Psi_{\rm Kaons}(u)$ and $\Psi_{\rm Protons}(u)$ for pions, kaons and protons (see the upper panel of  \fref{fig:comp_u_pi_k_p_tsallis_diff}).  In order to illustrate this difference clearly, we present the distribution of $r=\Psi_{\rm Kaons (Protons)}(u)/\Psi_{\rm Pions}(u)$ in the lower panel of \fref{fig:comp_u_pi_k_p_tsallis_diff}. In the small (large) $p_{T}$ region with $u<$ 0.6 ($u>$ 2), $\Psi_{\rm Pions}(u)$ is larger than $\Psi_{\rm Protons}(u)$ and $\Psi_{\rm Kaons}(u)$, and the discrepancy goes down (up) with the increase of $u$. While in the moderate $p_{T}$ region with 0.6 $<u<$ 2, $\Psi_{\rm Pions}(u)$ is smaller than $\Psi_{\rm Protons}(u)$ and $\Psi_{\rm Kaons}(u)$. These differences could be validated by the comparison among the normalized moments of the transverse momentum distributions $\langle p_{T}^{n}\rangle/\langle p_{T}\rangle^{n}$ for pions, kaons and protons. This normalized moment is expressed in terms of $\Psi(u)$, $\langle p_{T}^{n}\rangle/\langle p_{T}\rangle^{n}=\int^{\infty}_{0}u^{n}\Psi(u)udu$. \Tref{tab:id_particles_normalized_moments} gives $\langle p_{T}^{n}\rangle/\langle p_{T}\rangle^{n}$ for the identified particles with $n=2,3,4$. Because of $\Psi_{\rm Pions}(u)$ being above $\Psi_{\rm Protons}(u)$ and $\Psi_{\rm Kaons}(u)$ at small (large) $u$, the normalized moment of the pion $p_{T}$ distribution is larger than the ones of the proton and kaon $p_{T}$ distributions for small (large) $n$. Identical results are obtained from the comparison between the normalized moments of pions and protons produced in Au+Au collisions at $\sqrt{s_{NN}}=$ 200 GeV \cite{proton_antiproton_spectra}.

\begin{table}[H]
\caption{\label{tab:id_particles_normalized_moments} The normalized moments of the transverse momentum distributions $\langle p_{T}^{n}\rangle/\langle p_{T}\rangle^{n}$  for pions, kaons and protons. The uncertainties quoted are due to the errors of $C^{'}_{q}$, $q^{'}$ and $u_{0}$ in \tref{tab:id_particles_normalized_parameters}.}
\begin{center}
\begin{tabular}{@{}cccc}
\hline  
\textrm{$n$}&
\textrm{Pions}&
\textrm{Kaons}&
\textrm{Protons}
\\
\hline  
2 & 1.9$\pm$0.2 &1.7$\pm$0.1&1.5$\pm$0.2\\
3&6.5$\pm$1.0 &4.9$\pm$0.3&3.3$\pm$0.5\\
4 &44.8$\pm$8.2 &26.8$\pm$2.2&10.7$\pm$2.2\\
\hline  
\end{tabular}
\end{center}
\end{table}

\section{Comparison to the scaling behaviour presented in $\tau$} \label{sec:comp}
\begin{figure}[h]
\centering
\includegraphics[scale=0.18]{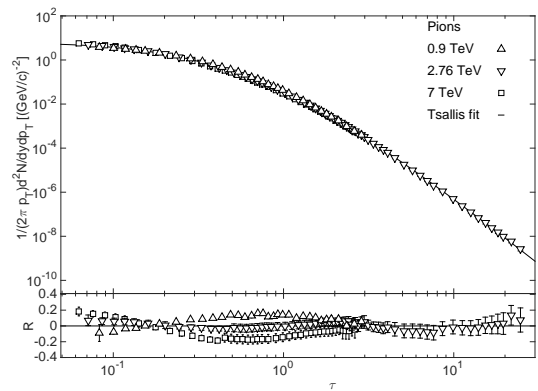} 
\caption{\label{fig:tau_p_p_scaling}Upper panel: the geometrical scaling of the pion $p_{T}$ spectra presented in $\tau$ for $pp$ collisions at 0.9, 2.76 and 7 TeV. The solid curve is described by the pion scaling function in equation \eref{eq:phi_z_pt_spectrum_pp} with $z$ replaced by $\tau$. The data points are taken from \cite{pt_spectra_0_9, pt_spectra_2_76, pt_spectra_7_0}. Lower Panel: the $R$ distribution as a function of $\tau$ for pions.}
\end{figure}

In \cite{geometrical_scaling}, a geometrical scaling behaviour appeared when the $p_{T}$ spectra of identified particles in $pp$ collisions at 0.9, 2.76 and 7 TeV were presented in terms of the scaling variable $\tau=p_{T}/Q_{0}(p_{T}/\sqrt{s})^{\lambda/2}$, where $Q_{0}=1$ GeV/c and $\lambda \approx$ 0.27. For the sake of illustrating this scaling behaviour, the author in that article chose the pion $p_{T}$ spectra at 7 TeV as a reference, and showed the ratio between the spectra at 7 and 0.9 (2.76) TeV as a function of $\tau$. In this work, in order to compare the scaling behaviours presented in $z$ and $\tau$ directly, we select the pion spectra at 2.76 TeV as a baseline, and plot the spectra at 0.9, 2.76 and 7 TeV as a function of $\tau$ in the upper panel of  \fref{fig:tau_p_p_scaling}.  The pion spectra at 2.76 TeV is described by the modified Tsallis fit in equation \eref{eq:phi_z_pt_spectrum_pp} with $z$ replaced by $\tau$. The lower panel of \fref{fig:tau_p_p_scaling} shows the $R$ distribution as a function of $\tau$. The data points at 0.9, 2.76 and 7 TeV agree with the fitted curve within an accuracy of 20$\%$. This accuracy is comparable to the one of the scaling behaviour presented in $z$.  Analogous results are obtained in the comparison between the scaling behaviours presented in $z$ and $\tau$ for the kaon and proton $p_{T}$ spectra. However, the author in \cite{geometrical_scaling} showed that when $\tau>$ 6, the geometrical scaling starts to break down. The data at 7 TeV utilized in that work are preliminary and have not been published by the ALICE collaboration so far.  In order to see whether there is a breaking in the scaling behaviour presented in $z$, more public data in the high $p_{T}$ region at 7 TeV are needed.

\section{Colour String Percolation Model}\label{sec:mechanism}
In  \cite{geometrical_scaling, gluon_saturation}, the author suggested the scaling behaviour presented in $\tau$ could be understood by the gluon saturation mechanism. However, in this work, we will argue that the scaling behaviour presented in $z$ for the identified particles could be explained by the colour string percolation (CSP) model \cite{string_perco_model_1, string_perco_model_2} in a quantitative way simultaneously.

\begin{figure}[h]
\centering
\includegraphics[scale=0.18]{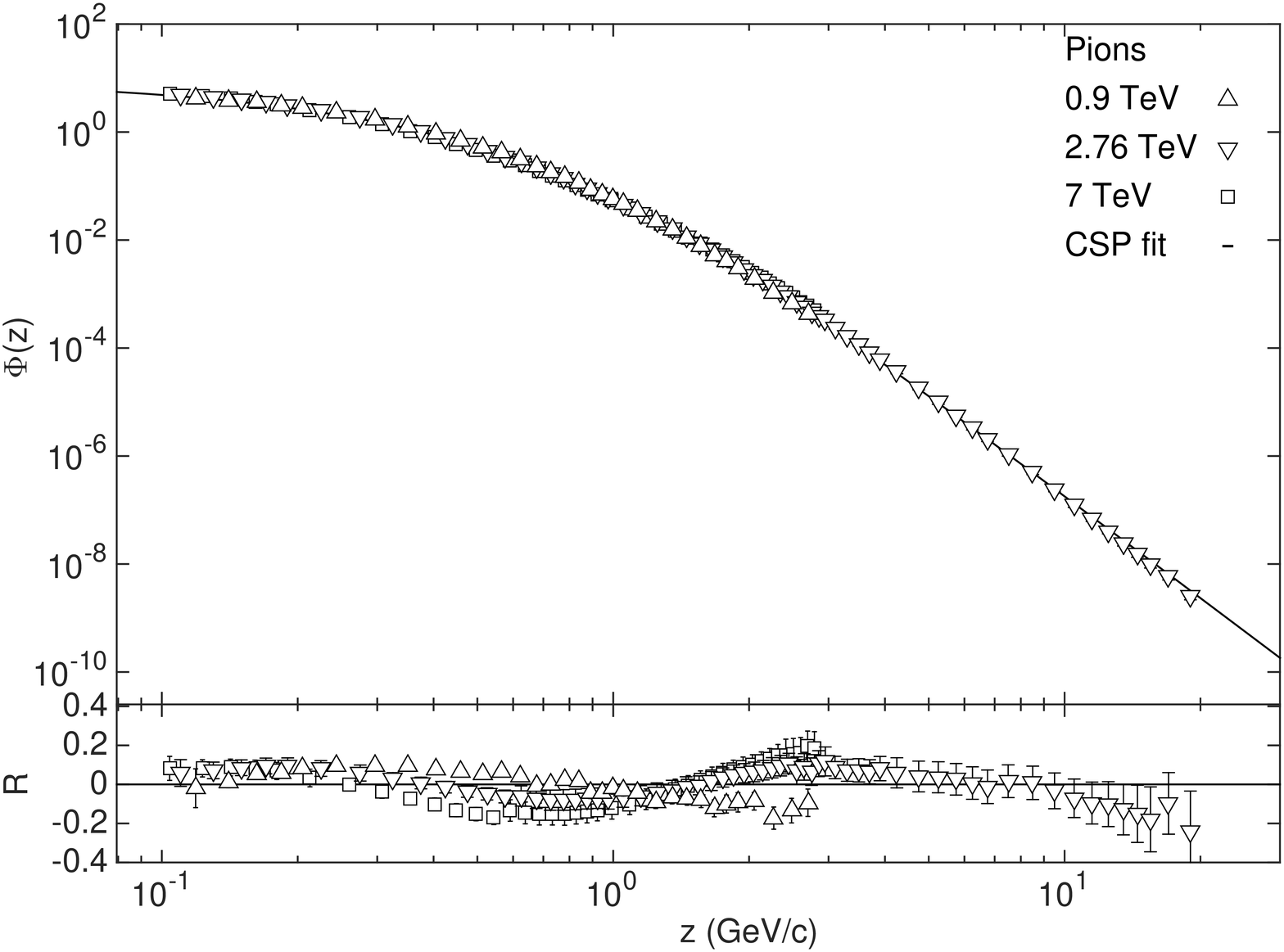} 
\caption{\label{fig:pi_csp_fit_plus_ratio_log_2_76_TeV_QF} Upper panel: the scaling behaviour of the pion $p_{T}$ spectra presented in $z$ for $pp$ collisions at  0.9, 2.76 and 7 TeV. The solid curve is described by the CSP fit in equation \eref{eq:CPS_formula} with parameters in the second column of \tref{tab:CSP_fit_parameters}. The data points are taken from \cite{pt_spectra_0_9, pt_spectra_2_76, pt_spectra_7_0}. Lower Panel: the $R$ distribution as a function of $z$ for pions.}
\end{figure}

In this model, colour strings are stretched between the partons of protons in $pp$ collisions. These strings then decay into new ones with the emission of $q\bar{q}$ pairs. Identified hadrons are produced through the hadronization of these new strings. The strings are viewed as small areas of $S_{1}=\pi r_{0}^{2}$ with $r_{0}\approx 0.2$ fm in the transverse plane. The number of strings grows with the increase of the collision energy. When there are $n$ strings, they start to overlap and form clusters with transverse areas of $S_{n}$. The mean transverse momentum squared $\langle p_{T}^{2}\rangle_{ni}$ of identified particles produced by a cluster can be written as $\langle p_{T}^{2}\rangle_{ni}=\sqrt{nS_{1}/S_{n}} \langle p_{T}^{2}\rangle_{1i}$, where $\langle p_{T}^{2}\rangle_{1i}$ is the mean $p_{T}^{2}$ of identified particles produced by a single string, and $nS_{1}/S_{n}$ is the degree of string overlap. If strings just get in touch with each other, then $S_{n}=nS_{1}$ and $nS_{1}/S_{n}=1$. If strings maximally overlap with each other, then $S_{n}=S_{1}$ and $nS_{1}/S_{n}=n$ with $n>1$. The $p_{T}$ spectra of identified particles produced in $pp$ collisions is deemed as a superposition of the $p_{T}$ spectra produced by each cluster, $f(x, p_{T})$, weighted with the distribution of the cluster's size, $W(x)$. Here the cluster size $x$ is equivalent to the inverse of  $\langle p_{T}^{2}\rangle_{ni}$. In \cite{string_perco_model_1} and \cite{string_perco_model_2}, the fragmentation function for the cluster is chosen as the Schwinger formula $f(x, p_{T})=\textrm {exp}(-p_{T}^{2}x)$ \cite{schwinger_formula}. However, this function only describes the spectra well at the soft $p_{T}$ region. In order to describe the spectra at the high $p_{T}$ region as well, the Schwinger formula should be replaced by $f(x, p_{T})=\textrm {exp}(-\sqrt{2x}p_{T})$ \cite{frag_function_new_1, frag_function_new_2}. $W(x)$ is supposed to be a gamma distribution, $W(x)=\frac{\gamma}{\Gamma(k)}(\gamma x)^{k-1}\textrm{exp}(-\gamma x)$, where $\gamma$ and $k$ are free parameters. Thus the $p_{T}$ distribution of identified particles in $pp$ collisions is 
\begin{eqnarray}
\frac{d^{2}N}{2\pi p_{T}dp_{T}dy}=C\int_{0}^{\infty}W(x)f(x,p_{T})dx,
\label{eq:CPS_formula}
\end{eqnarray}
where $C$ is a normalization constant for the total number of clusters formed for identified particles before hadronization. In order to see whether the CSP model can describe the scaling behaviour presented in $z$, we fit equation \eref{eq:CPS_formula} to the $p_{T}$ spectra of identified particles at 2.76 TeV. The parameters $C$, $r$ and $k$ returned by the fits as well as the reduced  $\chi^{2}$s are tabulated in  \tref{tab:CSP_fit_parameters}. As shown in the upper panel of \fref{fig:pi_csp_fit_plus_ratio_log_2_76_TeV_QF}, the data points for pions agree with the CSP fit well in log scale. The accuracy of this agreement is around $20\%$, which can be seen from the $R$ distribution in the lower panel of the figure. Similar results are obtained for kaons and protons. As the $k$ parameters are different for the identified particles, pions, kaons and protons originate from different distributions of clusters. 

\begin{table}[H]
\caption{\label{tab:CSP_fit_parameters} The parameters $C$, $r$ and $k$ of the CSP fits on the spectra of pions, kaons and protons at 2.76 TeV. The uncertainties quoted are due to the statistical plus systematic errors of the data points added in quadrature. The last line shows the reduced $\chi^{2}$s for the fits.}
\begin{center}
\begin{tabular}{@{}cccc}
\hline  
\textrm{\ }&
\textrm{Pions}&
\textrm{Kaons}&
\textrm{Protons}
\\
\hline  
$C$ & 9.1$\pm$0.2 & 0.53$\pm$0.01& 0.160$\pm$0.007\\
$\gamma$&0.140$\pm$0.003& 0.342$\pm$0.008&0.91$\pm$0.07\\
$k$ & 3.15$\pm$0.02 & 3.33$\pm$0.02&5.0$\pm$0.2\\
$\chi^{2}$/dof& 1.53 & 0.91& 1.91\\
\hline  
\end{tabular}
\end{center}
\end{table}

Now we would like to seek for the reason why the CSP model can describe the scaling behaviour of the identified $p_{T}$ spectra. Under the transformation $x \rightarrow x' = \lambda x$, $\gamma \rightarrow \gamma' = \gamma/\lambda$ and $p_{T} \rightarrow p_{T}' = p_{T}/\sqrt{\lambda}$, both $W(x)$ and $f(x)$ are invariant. So the $p_{T}$ spectra of identified particles in equation \eref{eq:CPS_formula} are also invariant. This invariance is the scaling behaviour we are looking for. Comparing the $p_{T}'$ transformation in the CSP model $p_{T}' \rightarrow p_{T}'\sqrt{\lambda}$ with the one used to search for the scaling behaviour $p_{T}\rightarrow p_{T}/K$, we know that $K$ is proportional to $1/\sqrt{\lambda}$. As described in \cite{string_perco_model_2,frag_function_new_2}, $\lambda=\langle S_{n}/nS_{1} \rangle^{1/2}$, where the average is taken over all clusters for identified particles, thus $K$ should be proportional to $\langle nS_{1}/S_{n} \rangle^{1/4}$. Since the degree of string overlap $nS_{1}/S_{n}$ grows with energy, the scaling parameter $K$ should also increase with energy. That's indeed what we see for the identified particles in the third column of \tref{tab:id_particles_a_k_parameters}. Therefore the CSP model can explain the scaling behaviours presented in $z$ for pions, kaons and protons separately. However, the rate of the $K$ values increasing with energy for pions is different with the one for kaons or protons. This could also be explained by the CSP model. As $K= \langle p_{T}\rangle/\langle z \rangle$, and the values of $\langle z \rangle$ are the same for pions (kaons or protons) at 0.9, 2.76 and 7 TeV, the ratio between the values of $K$ at different energies should be equal to the ratio between the values of $\langle p_{T}\rangle$ at different energies. $\langle p_{T}\rangle$ is evaluated in terms of the CSP model as \cite{geometrical_scaling}
\begin{eqnarray}
\langle p_{T}\rangle=\frac{\int_{0}^{\infty}\int_{0}^{\infty}W(x)f(x,p_{T})p_{T}^{2}dxdp_{T}}{\int_{0}^{\infty}\int_{0}^{\infty}W(x)f(x,p_{T})p_{T}dxdp_{T}}.
\label{eq:p_T_CSP}
\end{eqnarray}
The cluster fragmentation functions $f(x,p_{T})$ are the same for pions (kaons or protons) at different energies. However, the cluster's size distributions $W(x)$ are different for different particles, which can be seen from the comparison among the $k$ values for pions, kaons and protons in \tref{tab:CSP_fit_parameters}. As a result, for different species of particles, the values of $K$ change in a different rate with energy. In a summary, the CSP model could be utilized to explain the scaling behaviours presented in $z$ for the identified particles simultaneously.

\section{Conclusions}\label{sec:conclusion}

In this paper, we have extended the scaling behaviour in the inclusive charged hadron $p_{T}$ spectra to the spectra of identified particles at 0.9, 2.76 and 7 TeV. This scaling behaviour is exhibited when $p_{T}$ is replaced by $z=p_{T}/K$. The scaling parameter $K$ is determined by the quality factor method, which does not rely on the shape of the scaling function. We have argued that the pions, kaons and protons originate from different distributions of clusters formed by strings overlapping, and the scaling behaviours of these identified particles could be explained by the colour string percolation model in a quantitative way at the same time.

\section*{Acknowledgements}
The author would like to thank Prof. C. B. Yang at Central China Normal University for valuable discussions. This work was supported by the Fundamental Research Funds for the Central Universities of China under Grant No. GK201502006, by the Scientific Research Foundation for the Returned Overseas Chinese Scholars, State Education Ministry, and by the National Natural Science Foundation of China under Grant Nos. 11447024 and 11505108.


\end{document}